\documentclass{iaus}
\usepackage{graphicx}

\title[Stellar Evolution in the Early Universe] 
{Stellar Evolution in the Early Universe}

\author[Raphael Hirschi et~al.]   
{Raphael Hirschi$^{1,2}$,
Urs Frischknecht$^3$, F.-K. Thielemann$^3$, Marco Pignatari$^1$, Cristina Chiappini$^{4,5}$, Sylvia Ekstr\"om$^4$, Georges Meynet$^4$, \& Andr\'e Maeder$^4$}

\affiliation{$^1$Astrophysics group, Keele University, Lennard-Jones Lab., Keele, ST5 5BG, UK 
\\ email: {\tt r.hirschi@epsam.keele.ac.uk} \\[\affilskip]
$^2$IPMU, University of Tokyo, Kashiwa, Chiba 277-8582, Japan \\[\affilskip]
$^3$Dept. of Physics \& Astronomy, University of Basel, Klingelbergstr. 82, CH-4056, Basel, Switzerland\\[\affilskip]
$^4$Observatoire Astronomique de l'Universit\'e de Gen\`eve, CH-1290, Sauverny, Switzerland\\[\affilskip]
$^5$Osservatorio Astronomico di Trieste, Via G. B. Tiepolo 11, I - 34131 Trieste, Italia}

\pubyear{2008}
\volume{255}  
\pagerange{119--126}
\jname{Low-Metallicity Star Formation: From the First Stars to Dwarf Galaxies}
\editors{L.K. Hunt, S. Madden \& R. Schneider, eds.}

\newcommand{\el}[2]{\ensuremath{\rm^{#2}\kern-0.8pt\rm#1}}

\begin{document}

\maketitle

\begin{abstract}
Massive stars played a key role in the early evolution of the Universe. They formed
with the first halos and started the re-ionisation. It is therefore very important
to understand their evolution. In this paper, we describe the strong impact of
rotation induced mixing and mass loss at very low $Z$. The strong mixing leads to a
significant production of primary \el{N}{14}, \el{C}{13} and \el{Ne}{22}. Mass loss
during the red supergiant stage allows the production of Wolf-Rayet stars, type Ib,c
supernovae and possibly gamma-ray bursts (GRBs) down to almost $Z=0$ for stars more massive than
60 $M_\odot$. Galactic chemical evolution models calculated with models of rotating
stars better reproduce the early evolution of N/O, C/O and \el{C}{12}/\el{C}{13}.
We calculated the weak s-process production induced by the primary \el{Ne}{22} and 
obtain overproduction factors (relative to the initial composition, $Z=10^{-6}$) 
between 100-1000 in the mass range 60-90.
\end{abstract}

\firstsection 
\section{Introduction}
Massive stars ($M \gtrsim 10\ M_\odot$) started forming about 400 millions years after the Big Bang and ended the dark ages 
by re-ionising the Universe. They therefore played a key role in the early evolution of the Universe and
it is important to understand the properties and the evolution of the first stellar generations to
determine the feedback they had on the formation of the first cosmic structures. It is unfortunately not
possible to observe the first massive stars because they died a long time ago but their chemical signature 
can be observed in low mass halo stars (called EMP stars), which are so old and metal poor that 
the interstellar medium out of which these halo stars formed are thought to have been enriched
by one or a few generations of massive stars.
Since the re-ionisation, massive stars have continuously injected kinetic energy 
(via various types of supernovae) and 
newly produced chemical elements (by both hydrostatic and explosive burning and s and r processes) 
into the interstellar medium of their host galaxies. They are thus important players for the
chemo-dynamical evolution of galaxies. Most massive stars leave a remnant at their death, either a 
neutron star or a black hole, which often lead to the formation of pulsars or X-ray binaries.

The evolution of stars is governed by three main parameters, which are the initial mass, 
metallicity ($Z$) and 
rotation rate. The evolution is also influenced by the presence of magnetic fields and of a close binary
companion. For massive stars around solar metallicity, mass loss plays a crucial
role, in some cases removing more than half of the initial mass. Internal mixing, induced mainly by
convection and rotation also significantly affect the evolution of stars. 
The properties of non-rotating low-Z stars are summarised in Sect. 2 of Hirschi et~al. (2008). 
In short, at low Z, 
stars are more compact, usually have weaker winds (see however \cite{Pu08}) and are more massive 
(Bromm \& Loeb 2003, \cite{SOIF06}).
The fate of non-rotating massive single stars at low Z is summarised in \cite{HFWLH03}
and several groups have calculated the corresponding stellar yields (\cite{HW02,CL04,TUN07}).
In this paper, after discussing the impact of rotation 
induced mixing and mass loss at low $Z$, we present the implications for
the nucleosynthesis and for galactic chemical evolution in the context of extremely metal poor stars. 

\section{Rotation, internal mixing and mass loss}

Massive star models including the effects of both mass loss and especially rotation better 
reproduce many observables around solar $Z$. For example, models with rotation allow chemical surface enrichments
already on the main sequence (MS), whereas without the inclusion of rotation, self-enrichments are only possible
during the RSG stage (\cite{HL002,ROTV}). Rotating star models also better reproduce the 
WR to O type star
ratio and also the ratio of type Ib+Ic to type II supernova as a function of metallicity compared
to non-rotating models, which underestimate these ratios 
(see contribution by Georgy in this volume and \cite{ROTXI}). 
The models at very low $Z$ presented here
use the same physical ingredients as the successful solar $Z$ models.
The value of 300 km\,s$^{-1}$ used as the initial rotation velocity at solar 
metallicity
corresponds to an average velocity of about 220\,km\,s$^{-1}$ on the main
sequence (MS) which is
close to the average observed value. 
One of the first surveys of OB stars (and in particular their surface velocity) 
was obtained by \cite{FU82} and 
the most recent surveys are listed in Meynet et~al. (2008).
It is unfortunately not possible to observe very low $Z$ massive stars and measure their rotational 
velocity since they all died a long time ago.
Higher observed ratio of Be to B stars in
the Magellanic clouds compared to our Galaxy (\cite{MGM99}, \cite{MFH07}) could point
out to the fact the stars rotate faster at lower metallicities. 
Also a low-$Z$ star having the same ratio of surface velocity to critical
velocity, $\upsilon/\upsilon_{\rm crit}$ (where $\upsilon_{\rm crit}$ is the velocity for
which the centrifugal force balances the gravitational force)
as a solar-$Z$ star has a higher surface rotation velocity due to
its smaller radius (one quarter of $Z_\odot$ radius for a very low-$Z$ 20 $M_\odot$
star). In the models presented below, the initial ratio $\upsilon/\upsilon_{\rm crit}$
is the same or slightly higher than for solar Z (see \cite{H07} for more details). This
corresponds to initial surface velocities in the range of $600-800$ km\,s$^{-1}$. These
fast initial rotation velocities are supported by chemical evolution models of \cite{CH06} 
discussed in the next section.
The mass loss prescriptions used in the Geneva stellar evolution code are described in detail in
Meynet \& Maeder (2005). In particular, the mass loss rates depend 
on metallicity as $\dot{M} \sim (Z/Z_{\odot})^{0.5}$, where
$Z$ is the mass fraction of heavy elements at the surface
of the star.

How do rotation induced processes vary with metallicity? The surface layers of massive stars 
usually accelerate due to internal transport of angular momentum from the core to the envelope. 
Since at low $Z$, stellar winds are weak, this angular momentum dredged up by meridional
circulation remains in the star, and the star more easily reaches critical rotation. 
At the critical limit, matter can easily be launched into a keplerian disk which probably 
dissipates under the action of the strong radiation pressure of the star.

The efficiency of meridional
circulation (dominating the transport of angular momentum) decreases towards lower Z 
because the Gratton-\"Opik term of the vertical velocity of the outer cell is
proportional to $1/\rho$. On the other hand, shear mixing (dominating the mixing of chemical elements) 
is more efficient at low $Z$. Indeed, the star is
more compact and therefore the gradients of angular velocity are larger and the mixing 
timescale (proportional to the square of the radius) is shorter. This leads to stronger
internal mixing of chemical elements at low Z (\cite{ROTVIII}).

The history of convective zones (in particular the convective zones
associated with shell H burning and core He burning) is strongly affected by rotation
induced mixing (see \cite{H07}). 
The most important rotation induced mixing takes place while helium is burning inside a convective core.
Primary carbon and oxygen are
mixed outside of the convective core into the H-burning shell. Once the
enrichment is strong enough, the H-burning shell is boosted (the CNO
cycle depends strongly on the carbon and oxygen mixing at such low
initial metallicities). The shell then becomes convective and leads to an important
primary nitrogen production.
In response to the shell boost, the core
expands and the convective core mass decreases.
At the end of He burning, the CO core is less massive than in 
the non-rotating model. 
Additional convective and rotational
mixing brings the primary CNO to the surface of the star.
This has consequences for the stellar yields.
The yield of $^{16}$O, being closely
correlated with the mass of the CO core, is therefore
reduced due to the strong mixing. 
At the same time the carbon yield is slightly increased.
The relatively ``low'' oxygen yields and ``high'' carbon
yields are produced over a large mass range at $Z=10^{-8}$ (\cite{H07}). This is
one possible explanation for the possible high [C/O] ratio observed in the most
metal-poor halo stars (see Fig. 14 in \cite{FS6} and \cite{FA08}) and in DLAs (\cite{PZ08}).

\begin{figure}[ht]
\begin{center}
\includegraphics[width=0.5\textwidth]{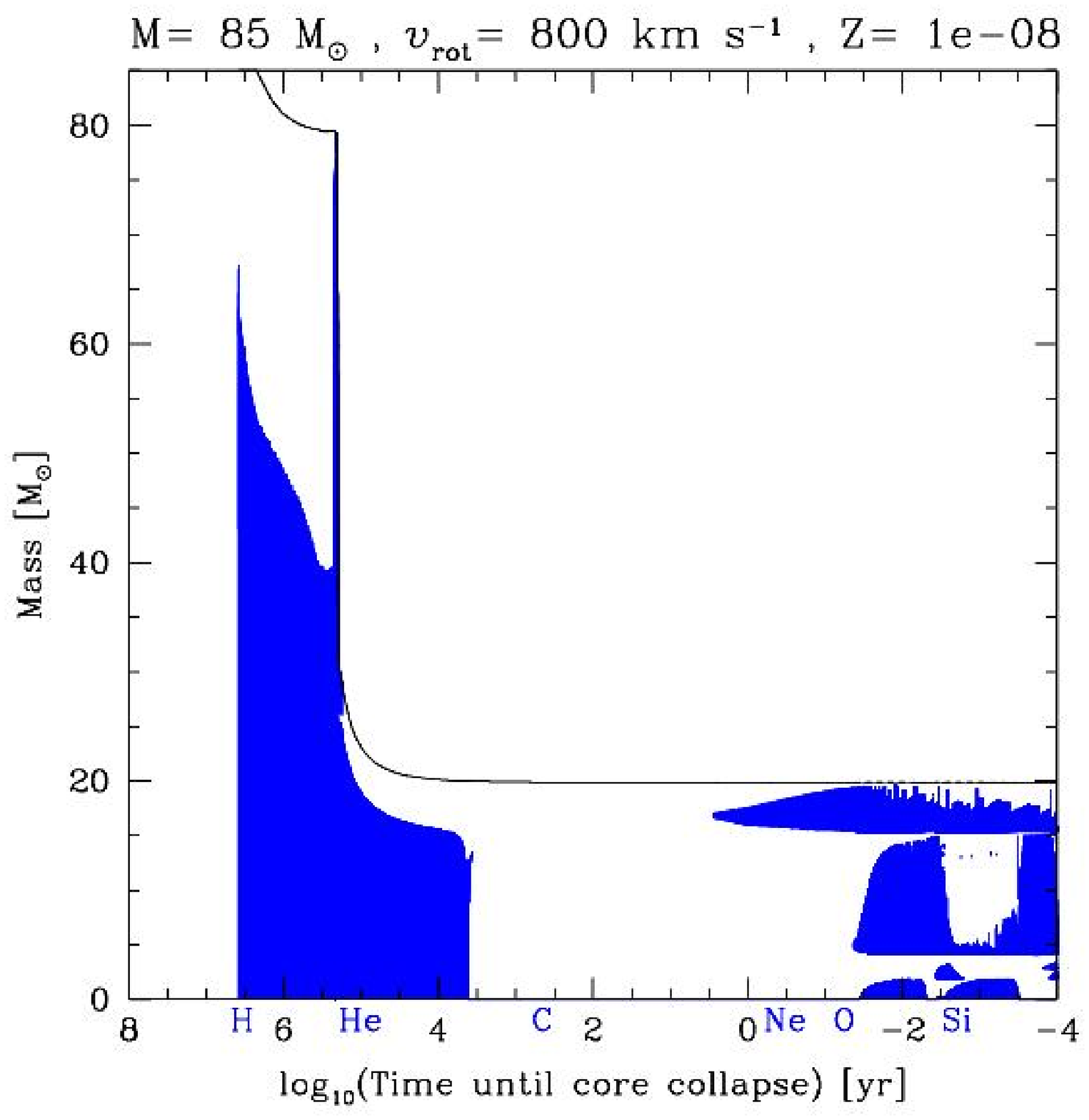}\includegraphics[width=0.5\textwidth]{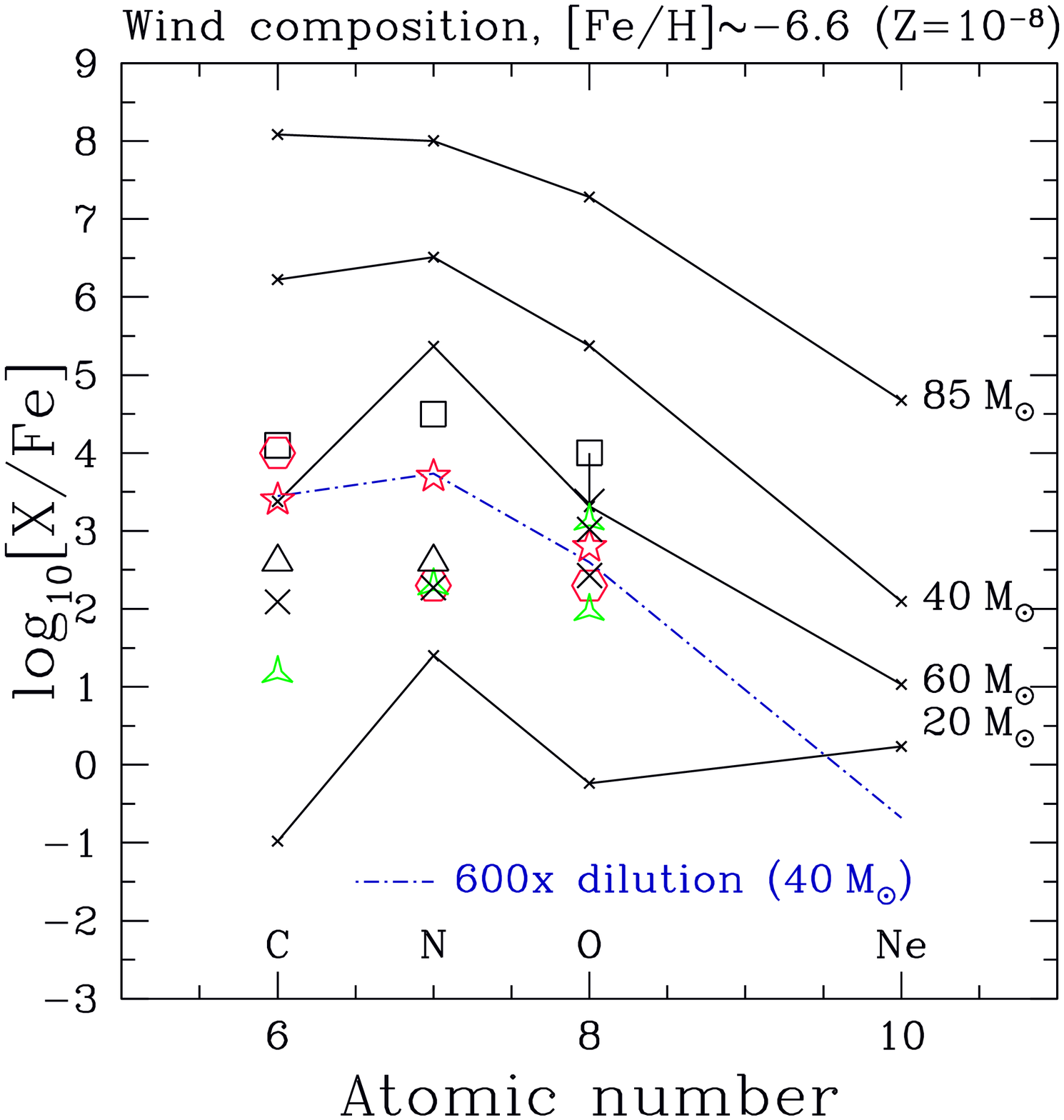}
\caption{{\it Left: }Structure evolution diagram for a 85 $M_\odot$ model 
with $\upsilon_{\rm ini}=800$ km\,s$^{-1}$ 
at $Z=10^{-8}$. Coloured areas correspond to convective zones along the Lagrangian mass 
coordinate as a function of the time left until the core collapse. 
The burning stage abbreviations are given below the time axis. The top solid line shows the total mass of the
star. A strong mass loss during the RSG stage removes a large fraction of total
mass of the star.
{\it Right: }Composition in [X/Fe] of the stellar wind for the $Z=10^{-8}$
models (solid lines).
For HE1327-2326 ({\it red stars}), the best fit for the
CNO elements is
obtained by diluting the composition of the wind of the 40 $M_\odot$
model by a factor 600 (see \cite{H07} for more details).}
\label{kip85-cemp}
\end{center}
\end{figure}

Models of metal-free stars including the effect of rotation 
(See contribution by Ekstr\"om and \cite{EM08})
show that stars may lose up to
10 \% of their initial mass due to the star rotating at its critical limit (also
called break-up limit). 
The mass loss due to the star reaching the critical
limit is non-negligible but not important enough to change
drastically the fate of the metal-free stars. 
The situation is very different at very low but non-zero metallicity 
(\cite{MEM06,H07}). 
The total mass of an 85\,$M_\odot$ model at $Z=10^{-8}$ is shown in Fig.
\ref{kip85-cemp} ({\it left}) by the top solid line. This model, like metal-free models,
loses around 5\% of its initial mass when its surface reaches break-up velocities in
the second part of the MS. At the end of core H burning,
the core contracts and the envelope expands, thus decreasing the surface
velocity and its ratio to the critical velocity. The mass loss rate becomes
very low again until the star crosses the HR diagram and reaches the RSG
stage. In the cooler part of the H-R diagram, the mass loss becomes very important. This is due to the dredge-up by the convective envelope of CNO elements to
the surface increasing its overall metallicity. The total metallicity, $Z$, is 
used in this model (including CNO elements)
for the metallicity dependence of the mass loss.
Therefore depending on how much CNO is brought up to the surface, the
mass loss becomes very large again. The CNO brought to the surface
comes from primary C and O produced in the He-burning region and from primary N produced
in the H-burning one.


The fate of rotating stars at very low Z is therefore probably the following:
\begin{itemize}
\item $M < 40 \ M_{\odot}$: Mass loss is insignificant and matter is only ejected
into the ISM during the SN explosion (see contributions by
Nomoto and Tominaga in this volume), which could be very energetic if
fast rotation is still present in the core at the core collapse.
\item $40\ M_\odot < M < 60\ M_\odot$: Mass loss (at critical rotation and
in the RSG stage) removes 10-20\% of the initial mass of the star. The star
probably dies as a black hole without a SN explosion and therefore the
feedback into the ISM is only due to stellar winds, which are slow.
\item $M > 60\ M_\odot$: A strong mass loss removes a significant amount of
mass and the stars enter the WR phase. These stars therefore die as type Ib/c
SNe and possibly as GRBs.
\end{itemize}

\section{Nucleosynthesis and galactic chemical evolution}
Rotation induced mixing leads to the production of primary nitrogen,
\el{C}{13} and \el{Ne}{22}. In this section, we compare the
chemical composition of our models with carbon-rich EMP stars and 
include our stellar yields in a galactic chemical evolution (GCE) model and 
compare the GCE model with observations of EMP stars. We also study the weak s process 
production that can be expected with the primary \el{Ne}{22} obtained in our models.

\subsection{The most metal-poor star known to date, HE1327-2326}

Significant mass loss in very low-$Z$ massive stars offers an interesting
explanation for the strong enrichment in CNO elements of the most 
metal-poor stars observed in the halo of the galaxy 
(see \cite{MEM06,H07}). 
The most metal-poor star known to date, 
HE1327-2326 (\cite{Fr06}) is characterised by very high N, C and O abundances,
high Na, Mg and Al abundances, a weak s-process enrichment and depleted
lithium. The star is not evolved so it has not had time to bring
self-produced CNO elements to its surface and is most likely a subgiant.
By using one or a few SNe and using a very large mass cut, 
\cite{LCB03} and \cite{IUTNM05} are
able to reproduce the abundance of most elements. 
However they are not
able to reproduce the nitrogen surface abundance of
HE1327-2326 without rotational mixing. 
The abundance pattern observed at the surface of that star
presents many similarities with the abundance pattern obtained in the 
winds of very metal-poor fast rotating massive star models.
HE1327-2326 may therefore have formed
from gas, which was mainly enriched by stellar winds of rotating very low
metallicity stars. In this scenario, a first generation of stars 
(PopIII) 
pollutes the interstellar medium to very low metallicities
([Fe/H]$\sim$-6). Then a PopII.5 star 
(\cite{paris05}) like the 
40 $M_\odot$ model calculated here
pollutes (mainly through its wind) the interstellar medium out of
 which HE1327-2326 forms.
This would mean that HE1327-2326 is a third generation star.
The CNO abundances are well reproduced, in particular that of
nitrogen, which according to Frebel et~al. (2006)
is 0.9 dex higher in [X/Fe] than oxygen.
This is shown in Fig. \ref{kip85-cemp} ({\it right}) where the abundances of HE1327-2326 are
represented by the red stars and the best fit is 
obtained by diluting the composition of the wind of the 40 $M_\odot$
model by a factor 600. When the SN
contribution is added, the [X/Fe] ratio is usually lower for nitrogen
than for oxygen. 
%
It is interesting to note that the very high CNO yields of the 
40 $M_\odot$ stars brings the total
metallicity $Z$ above the limit for low mass star formation
obtained in \cite{BL03}.

\subsection{Primary nitrogen and \el{C}{13}}

\begin{figure}
\begin{center}
  \includegraphics[width=0.7\textwidth]{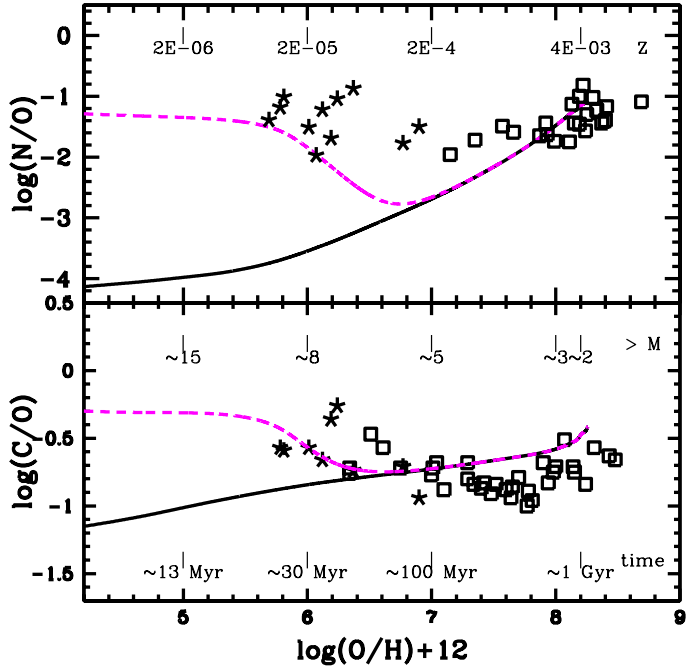}
  \caption{Chemical evolution model predictions of the N/O and C/O
evolution, in the galactic
halo, for different stellar evolution inputs. 
The solid curves show the predictions of a model without
fast rotators at low metallicities. The dashed lines show the effect 
of including a population of fast
rotators at low metallicities (for the data see \cite{CH06b} and references therein).}
\label{CNO}
\end{center}
\end{figure}

The high N/O plateau values observed at the surface of very metal poor 
halo stars
require very efficient sources of primary nitrogen. Rotating massive stars
can inject in a short timescale large amount of primary N. They are 
therefore very
good candidates to explain the N/O plateau observed at very low metallicity.
According to the heuristic model of \cite{CMB05}, a primary nitrogen production 
of about 0.15 $M_\odot$ per star is necessary.
Using stellar yields calculations taking into account the effects of rotation at Z$=$10$^{-8}$
in a chemical evolution model for the galactic halo with infall and outflow, 
both high N/O and C/O ratios are obtained in the very metal-poor
metallicity range in agreement with observations 
(see details in \cite{CH06b}). This model is shown in Fig. \ref{CNO}
(dashed magenta curve). In the same figure, a model computed without
fast rotators (solid black curve) is also shown. Fast rotation
enhances the nitrogen production by $\sim$3 orders of magnitude.
These results also offer a natural explanation for the large 
scatter observed in the N/O abundance ratio of {\it normal} metal-poor 
halo stars: given the strong dependency of the nitrogen yields on the 
rotational velocity of the star, we expect a
scatter in the N/O ratio which could be the consequence of the distribution 
of the stellar rotational velocities as a function of metallicity. 

As explained above, the strong production of primary nitrogen
is linked to a very active H-burning shell and therefore a smaller helium core.
As a consequence, less carbon is turned 
into oxygen, producing high C/O ratios (see Fig. 2).
Although the abundance data for C/O is still very uncertain, 
a C/O upturn at low
metallicities is strongly suggested by observations (see \cite{FA08}).
Note that this upturn is now also observed in very metal poor DLA
systems (\cite{PZ08}).

In addition, stellar models of fast
rotators have a great impact on the evolution of the
\el{C}{12}/\el{C}{13} ratio at very low metallicities (Chiappini et~al. 2008).
In this case, we predict that, if fast rotating massive stars were
common phenomena in the early Universe, the primordial interstellar
medium of galaxies with a star formation history similar to the one
inferred for our galactic halo should have \el{C}{12}/\el{C}{13}
ratios between 30-300. Without fast rotators, the predicted
\el{C}{12}/\el{C}{13} ratios would be $\sim 4500$ at [Fe/H] $= -3.5$,
increasing to $\sim 31000$ at around [Fe/H] $= -5.0$ (see Fig.2 in Chiappini et~al. 2008).
Current data on EMP giant normal stars in the galactic halo
(\cite{FSIX}) agree better with chemical evolution models including
fast rotators. The expected difference in the \el{C}{12}/\el{C}{13}
ratios, after accounting for the effects of the first dredge-up,
 between our predictions
with/without fast rotators is of the order of a factor of 2-3.  However, larger
differences (a factor of $\sim 60-90$) are expected for giants at [Fe/H]$=-5$ or
turnoff stars already at [Fe/H]$=-3.5$. To test our predictions, challenging
measurements of the \el{C}{12}/\el{C}{13} in more extremely metal-poor giants and
turnoff stars are required.

\begin{figure}[htb]
\begin{center}
 \includegraphics[width=0.5\textwidth]{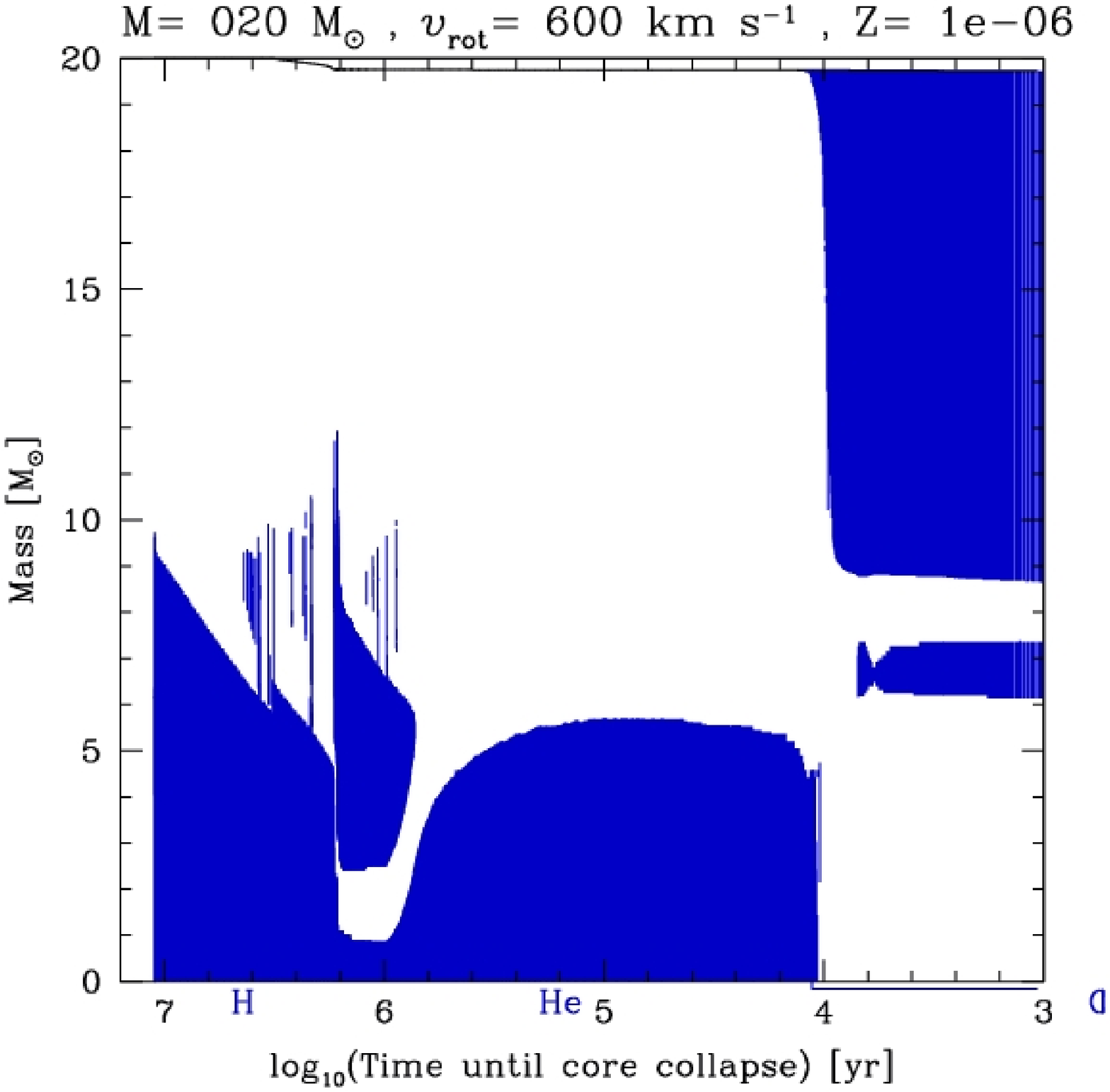}\includegraphics[width=0.5\textwidth]{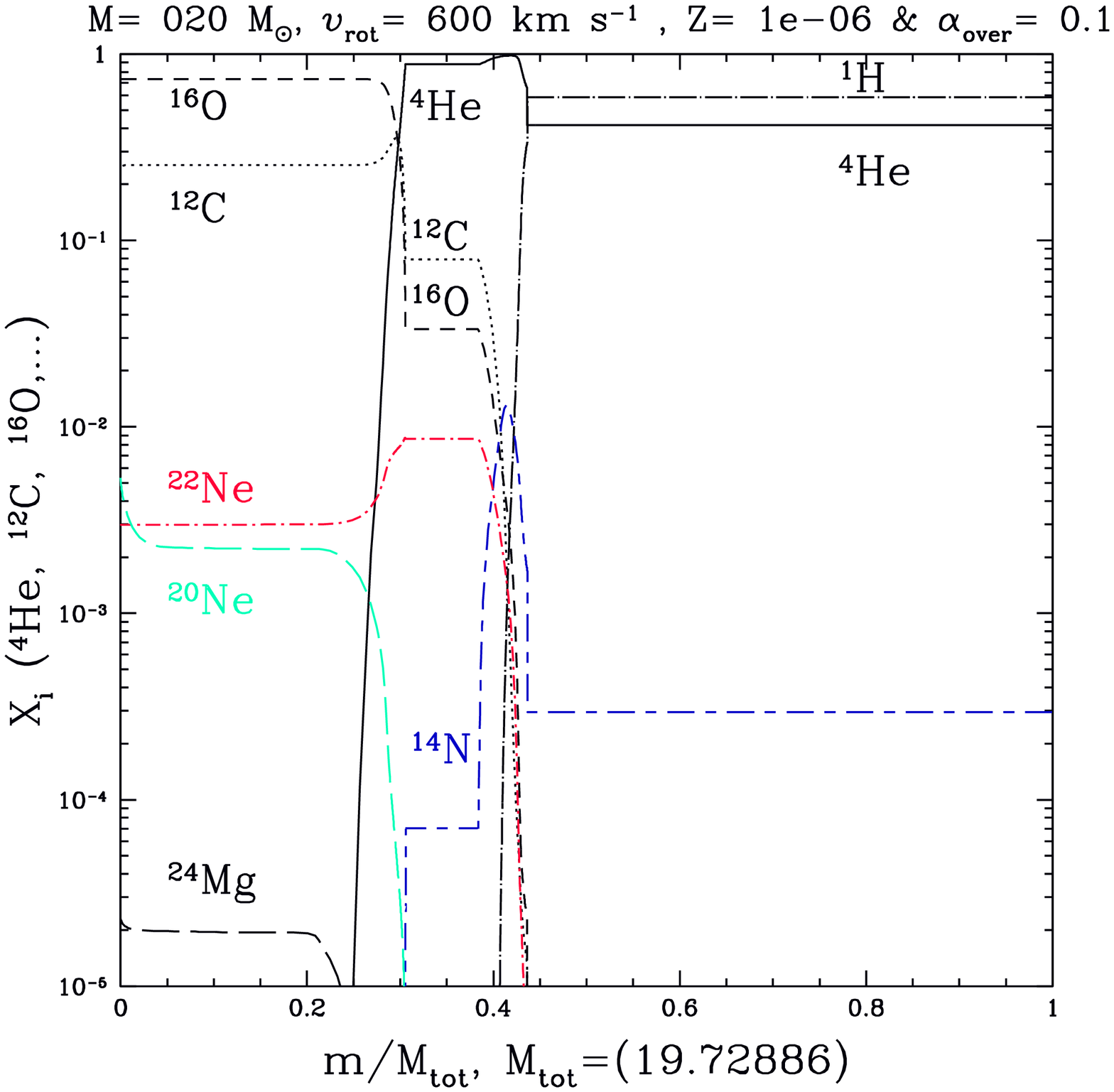} 
 \caption{{\it Left}: Structure evolution diagram (see description of Fig. \ref{kip85-cemp}) 
for a rotating 20 $M_\odot$ model 
at $Z=10^{-6}$ during H- and He-burning phases.
{\it Right}: Chemical composition at the end of core He burning. Just above the core,
one sees that the maximum abundance of \el{Ne}{22} is around 1\% in mass fraction and
at the end of core He burning, around 0.5\% is burnt in the core, providing
plenty of neutrons for s process.
}
\label{ne22}
\end{center}
\end{figure}

\subsection{Primary \el{Ne}{22} and $s$ process at low Z}
\begin{figure}[htb]
\begin{center}
\includegraphics[width=1.0\textwidth]{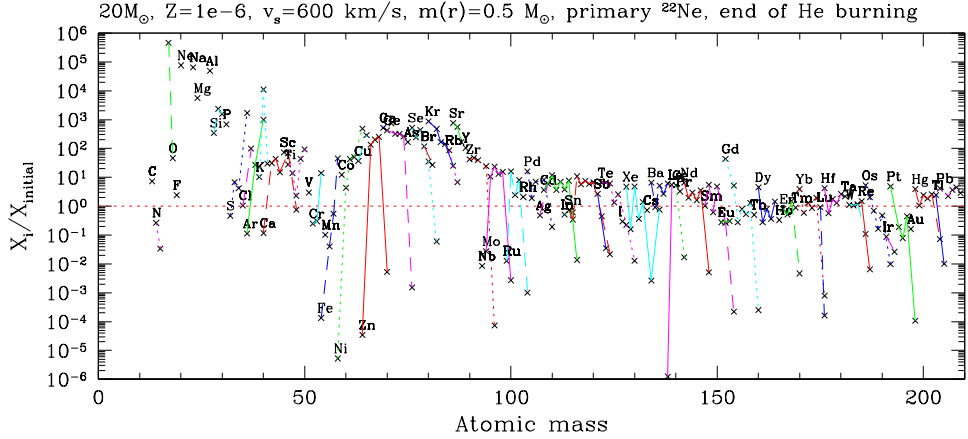}
 \caption{{\it Left}: Ratio of the abundance at the end of He-burning to the initial abundance 
 for a model including primary \el{Ne}{22}. We obtain large overproduction factors between 100 and 1000
 in the mass range (A) 60 to 90.
}
\label{sproc}
\end{center}
\end{figure}

Models at $Z=10^{-8}$ show a production of primary \el{Ne}{22} during He burning. 
We also started
calculating models at different Z to determine over which Z range the primary production of \el{Ne}{22}
and also \el{N}{14} is important. In Fig. \ref{ne22}, we show the properties of a 
20 $M_\odot$ model at $Z=10^{-6}$ up to the end of He burning. Around 0.5\% (in mass
fraction) of \el{Ne}{22} is burnt during core He burning and therefore leads to a
significant neutron release. 
We calculated the s process by using the primary nitrogen of the rotating 
20 $M_\odot$ model at $Z=10^{-6}$ inside a one-zone model (based on the Basel network and 
using an updated version of the {\it reaclib} reaction rate library) for s-process during He-burning. The first results are shown in Fig. \ref{sproc}.
Large overproduction factors (100-1000) are obtained, however the process is not primary 
(see Pignatari et~al. 2008 for more details).

\section{Conclusions and outlook}
The inclusion of
the effects of rotation changes significantly the simple picture in which 
stellar evolution at low Z is just stellar evolution without mass
loss. 
A strong mixing is
induced between the helium and hydrogen burning layers leading to a significant
production of primary \el{N}{14}, \el{C}{13} and \el{Ne}{22}. Rotating stellar models 
also predict a
strong mass loss during the RSG stage for stars more massive than 60 $M_\odot$. These models predict the formation of WR and
type Ib/c SNe down to almost Z=0.
The chemical composition of the stellar winds is compatible 
with the CNO abundance observed in the most metal-poor star known to date, HE1327-2326.
GCE models including the stellar yields of these rotating star models are able to better
reproduce the early evolution of N/O, C/O and \el{C}{12}/\el{C}{13} in our galaxy. These
new stellar evolution models predict a large neutron release during core He burning and 
we present here the corresponding s-process production during He burning. 

Large surveys of EMP stars (SEGUE, OZ surveys), of GRBs and SNe 
(Swift and GLAST satellites) and of massive stars (e.~g. VLT FLAMES survey) are underway 
and will bring more information and constraints on the evolution of massive stars at
low Z. On the theoretical side, more models are necessary to fully understand and study the complex interplay
between rotation, magnetic fields, mass loss and binary interactions at different
metallicities. Large grids of models at low Z will have many applications, 
for example to study the evolution of massive stars and their feedback in
high redshift objects like Lyman-break galaxies and damped Ly-alpha systems.



\section{Acknowledgements} R. Hirschi acknowledges financial support from 
the Royal Society (Conference Grant round 2008/R1) and from the organizers.

\end{document}